\documentclass[journal]{IEEEtran}
\usepackage{graphicx}
\usepackage[tight,footnotesize]{subfigure}
\usepackage[cmex10]{amsmath}
\usepackage{amsfonts}
\usepackage{amssymb}
\usepackage{textcomp}
\usepackage{bbm}        
\usepackage{algorithm}
\usepackage{algorithmic}
\usepackage{booktabs}   
\usepackage{multirow}
\usepackage{ifthen}
\usepackage{robinmath}

\usepackage{cite}

\ifCLASSINFOpdf
\else
\fi
\usepackage{array}

\usepackage{mdwmath}
\usepackage{mdwtab}

\usepackage{stfloats}
\usepackage{url}

\begin{document}
\hyphenation{two-di-men-sion-al}
\title{Fast Continuous Haar and Fourier Transforms of Rectilinear Polygons from VLSI Layouts}
%
%
%

\author{Robin~Scheibler,~\IEEEmembership{Member,~IEEE,}
        Paul~Hurley,~\IEEEmembership{Senior Member,~IEEE,}
        Amina~Chebira~\IEEEmembership{Member,~IEEE}%
\thanks{R. Scheibler and P. Hurley are with the System Software group at IBM Research --
        Z\"urich, S\"aumerstrasse 4, 8803 R\"uschlikon, Switzerland
        (robin.scheibler@ieee.org and pah@zurich.ibm.com).}%
\thanks{A. Chebira is with the Audiovisual Communications Laboratory (LCAV),
        School of Computer \& Communication Sciences,
        \'Ecole Polytechnique F\'ed\'erale de Lausanne (EPFL),
        1015 Lausanne, Switzerland (amina.chebira@epfl.ch).}}

\maketitle


\begin{abstract}
We develop the pruned continuous Haar transform and the fast continuous Fourier
series, two fast and efficient algorithms for rectilinear polygons. Rectilinear
polygons are used in VLSI processes to describe design and mask layouts of
integrated circuits. The Fourier representation is at the heart of many of
these processes and the Haar transform is expected to play a major role in
techniques envisioned to speed up VLSI design. To ensure correct printing of
the constantly shrinking transistors and simultaneously handle their
increasingly large number, ever more computationally intensive techniques are
needed. Therefore, efficient algorithms for the Haar and Fourier transforms are
vital.

We derive the complexity of both algorithms and compare it to that of discrete
transforms traditionally used in VLSI. We find a significant reduction in
complexity when the number of vertices of the polygons is small, as is the case
in VLSI layouts. This analysis is completed by an implementation and a
benchmark of the continuous algorithms and their discrete counterpart. We show
that on tested VLSI layouts the pruned continuous Haar transform is 5 to 25
times faster, while the fast continuous Fourier series is 1.5 to 3 times
faster.
\end{abstract}


%
\IEEEpeerreviewmaketitle

\newcommand{\NA}{\text{{\it NA}}}

\section{Introduction}
%
%
%
%


%

\IEEEPARstart{M}{oore's} famous law from 1965~\cite{moore_cramming_1965}
has been a major driving force in the effort to shrink transistors in Very
Large Scale Integration (VLSI). The tremendous progress in optical lithography
has been the enabler for massive production of integrated circuits and a
dramatic decrease in unit cost.

In optical lithography \cite{mack_fundamental_2008}, patterns of the integrated
circuits are transferred to silicon by shining light through a mask
and subsequently using a lens to concentrate the light onto a photosensitive
layer, as depicted in \ffref{lithobasic}. This is followed by an etching step,
which transfers the pattern to the silicon.
The Fourier transform of the mask is extensively used in simulation of the
lithography process owing to Fourier transforming properties of lenses
\cite{goodman_introduction_2004}. The projected image is the convolution of the
mask with an ideal filter with frequency response $H(f,g) = 1$ if
$\sqrt{f^2+g^2} \leq \frac{\NA}{\lambda}$ and 0 otherwise, where $f$ and $g$
are the spatial frequencies, $\NA$ the numerical aperture of the lens and
$\lambda$ the wavelength of the source \cite{kwok-kit_wong_resolution_2001}.
The size of the smallest feature printable by straightforward means with such a
lithography system is $0.25\lambda/\NA$ \cite{schellenberg_resolution_2004}.

The smallest feature in integrated circuits, also called technology node, has
shrunk from a few micrometers in the early days of optical lithography to 32nm
for the latest commercially available technology. The next industry target is
to print 22nm with sufficient yield. One way to achieve this target is to
develop extreme ultra-violet light source \cite{stulen_extreme_1999}, which,
with a wavelength of 15nm, would enable reliable production of future
technology nodes. However, it will seemingly not be ready on time for the 22nm
node, and remains to be proved a viable technology. This means the current
state of the art light source with a wavelength of 193nm has to be used for the
22nm node. Along with a numerical aperture of $\NA=1.44$ in modern lithography
systems, this sets the smallest feature possible at 34nm.

For the 32nm technology node, the system has already been pushed past its apparent
limit. At 22nm, the system is operating far below its limit, resulting in
severe optical degradation due to printed shapes being much smaller than
allowed by the current wavelength of the light. As a consequence, ever more
burden is placed onto computationally intensive techniques to circumvent the
optical degradation, and thus ensure sufficient manufacturing yield. These
techniques, collectively known as \emph{computational lithography}, include
traditional resolution enhancement~\cite{kwok-kit_wong_resolution_2001},
source-mask optimization~\cite{rosenbluth_optimum_2001}, and inverse
lithography~\cite{poonawala_opc_2006}. They strive to exploit all degrees of
freedom in the lithography process, including illumination amplitude, direction
and phase \cite{schellenberg_resolution_2004}. All of these techniques rely on
computationally intensive simulation of the underlying physical processes. An
alternative is to replace the simulation altogether by a statistical model. For
example, Kryszczuk et al.~\cite{kryszczuk_direct_2010} avoid costly
physical simulation by directly predicting the printability of the layouts
using a classifier trained with feature vectors from orthogonal transforms.

VLSI layout file sizes are expanding rapidly, with an increasing number of
transistors packed into a single design. They can reach more than a terabyte
for a single layer in technology nodes under
40nm~\cite{international_technology_roadmap_for_semiconductors_2007_2007}.
This coincides unfortunately with the increasing complexity of the
aforementioned computational lithography algorithms. Taking these factors into
account, having highly efficient algorithms at various steps of the lithography
process becomes crucial.

VLSI layouts consist primarily of highly repetitive patterns of rectilinear
polygons, those containing only right angles. Their vertex description,
although very compact, is cumbersome to use for applications where a
fixed-length representation is required, in particular machine learning.
However, they are also very sparse in the Haar basis since the building blocks
of its basis functions are themselves rectangles. The Haar transform is thus a
potentially excellent candidate to provide fixed-length features for any
post-processing envisioned in computational lithography. To extract Haar
transform coefficients from such enormous amount of vertex descriptions of
polygons, one needs a fast algorithm.

The Fourier representation is crucial in computational lithography, especially
in simulation of the lithography process. The underlying physical processes is
continuous, and using the continuous Fourier series (CFS) would seem natural.
However, a discretization step, comprising the sampling of the polygons,
followed by a fast Fourier transform is instead standard practice.
Unfortunately, sampling introduces aliasing and thus reduces overall accuracy.
CFS does not suffer from aliasing. However, a fast and efficient algorithm is
required to make the use of CFS practical given the extreme amount of data to
process in VLSI layouts.

Despite its close affinity with rectilinear polygons, the Haar transform has
seen surprisingly little use in lithography-related algorithms. In 1985, Haslam
et al. used a discrete Haar transform to compress the Fourier precompensation
filters for electron-beam lithography \cite{haslam_two-dimensional_1985},
whereas Ma et al. used the Haar transform in inverse lithography to regularize
the obtained mask~\cite{ma_generalized_2007}. When only a sampled signal is
available, a continuous transform cannot be computed. Even the so-called
continuous wavelet transform is computed as the projection of a discrete signal
onto an overcomplete basis \cite{antoine_two-dimensional_2004}. For polygons it
is however possible to compute the inner products with continuous basis
functions. In \cite{scheibler_pruned_2010}, we introduced the pruned continuous
Haar transform (PCHT), a fast algorithm to compute the continuous Haar
transform coefficients. We further study this algorithm in more detail in this
paper.

\begin{figure}[t]
  \centering
    \includegraphics[width=0.60\linewidth]{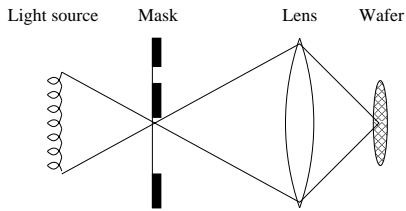}
  \caption{Basic principle of optical lithography. Patterns of the integrated
           circuit are transferred to the silicon wafer by shining light through a mask and
           subsequently using a lens to concentrate the light onto a photosensitive
           layer.}
  \flabel{lithobasic}
\end{figure}

As mentioned, the Fourier transform is extensively used at various
levels of the lithography process. Uses of the fast Fourier transform (FFT)
include the computation of a precompensation filter to reduce proximity effects
\cite{chow_image_1983}, and approximating the diffraction orders of the mask
\cite{rosenbluth_optimum_2001}. In \cite{chen_development_2008},
a road map for the efficient use of the 2D FFT in computational
lithography is developed. 
The mathematical idea to compute the CFS of polygons in itself is not new. In
1983, Lee and Mittra used geometry to derive a closed-form expression for the
Fourier integral over a polygon \cite{lee_fourier_1983}. A few years later, Chu
and Huang proposed a new and elegant derivation based on Stokes' theorem
\cite{chu_calculation_1989}. More recently, both Brandoli et al.
\cite{brandolini_average_1997} and Lu et al. \cite{lu_computable_2009} extended
the solutions to the $N$-dimensional case using the divergence theorem.

The inherently continuous nature of polygons makes a continuous transform a
natural tool for VLSI layouts. To the best of our knowledge, algorithms making
efficient use of this type of representation to quickly compute continuous
transform coefficients do not exist. We present in this paper the first fast
and efficient continuous Haar transform and Fourier series algorithms applied
to rectilinear polygons from VLSI layouts. They are based on a closed-form
formula that we derive for 2D continuous separable transforms using a
decomposition of rectilinear polygons into rectangles. This formulation allows
us to derive two fast algorithms to compute the transform coefficients: PCHT,
first introduced in \cite{scheibler_pruned_2010}, and the fast continuous
Fourier series (FCFS) algorithms. PCHT has a fast orthogonal wavelet transform
structure that is pruned using computational geometry techniques. FCFS results
from reducing the CFS computation problem to a few sparse discrete Fourier
transforms (DFT) computed using pruned FFT algorithms. We evaluate the
complexity of both algorithms and the performance of their implementation on
real VLSI layouts relative to their discrete counterparts. We find PCHT and
FCFS to be up to 25 and 3 times, respectively, faster than their discrete
counterparts.

This paper is organized as follows. \sref{background} introduces the necessary
background in VLSI layouts, the continuous Haar transform (CHT), and the CFS.
\sref{algo} presents a framework for taking continuous transforms of
rectilinear polygons as well as the proposed PCHT and FCFS. In
\sref{algo_perf_eval}, the performance of both algorithms is evaluated and
compared to that of their discrete counterparts. Finally, \sref{conclusion}
concludes by discussing the superiority of PCHT and FCFS over their discrete
equivalents, and sketches possible future directions.

%


\section{Background}
\slabel{background}

In this section, we first briefly describe VLSI layouts and how they are
created. The rectilinear polygons composing VLSI layouts are described
mathematically, laying the groundwork for the algorithms to come. We then
describe the 2D continuous Haar transform and the fast wavelet transform
algorithm used to compute it. Finally, a short refresher on the CFS is given.

\subsection{VLSI Layouts}

\subsubsection{Layouts}

VLSI layouts are composed of millions of rectangles, or more generally,
rectilinear polygons, which come from the transformation of elements of a
functional electric circuit. This data is then used for tasks such as mask
generation and printing using an optical lithography system.

Such a layout typically consists of several layers of various types.
\ffref{layout} shows fragments of three different types of layers. These are
taken from Metal 1 (M1), which is mostly random logic, Metal 2 (M2), containing
some logic and wires and Contact Array (CA), providing contacts between the
different layers. In addition, there are several other types of layers, omitted
here, similar to those of M1, M2 or CA.

\begin{figure}[tb]
  \centering
  \centerline{
    \begin{minipage}{0.25\linewidth}
      \centering
      \centerline{\includegraphics[width=\linewidth]{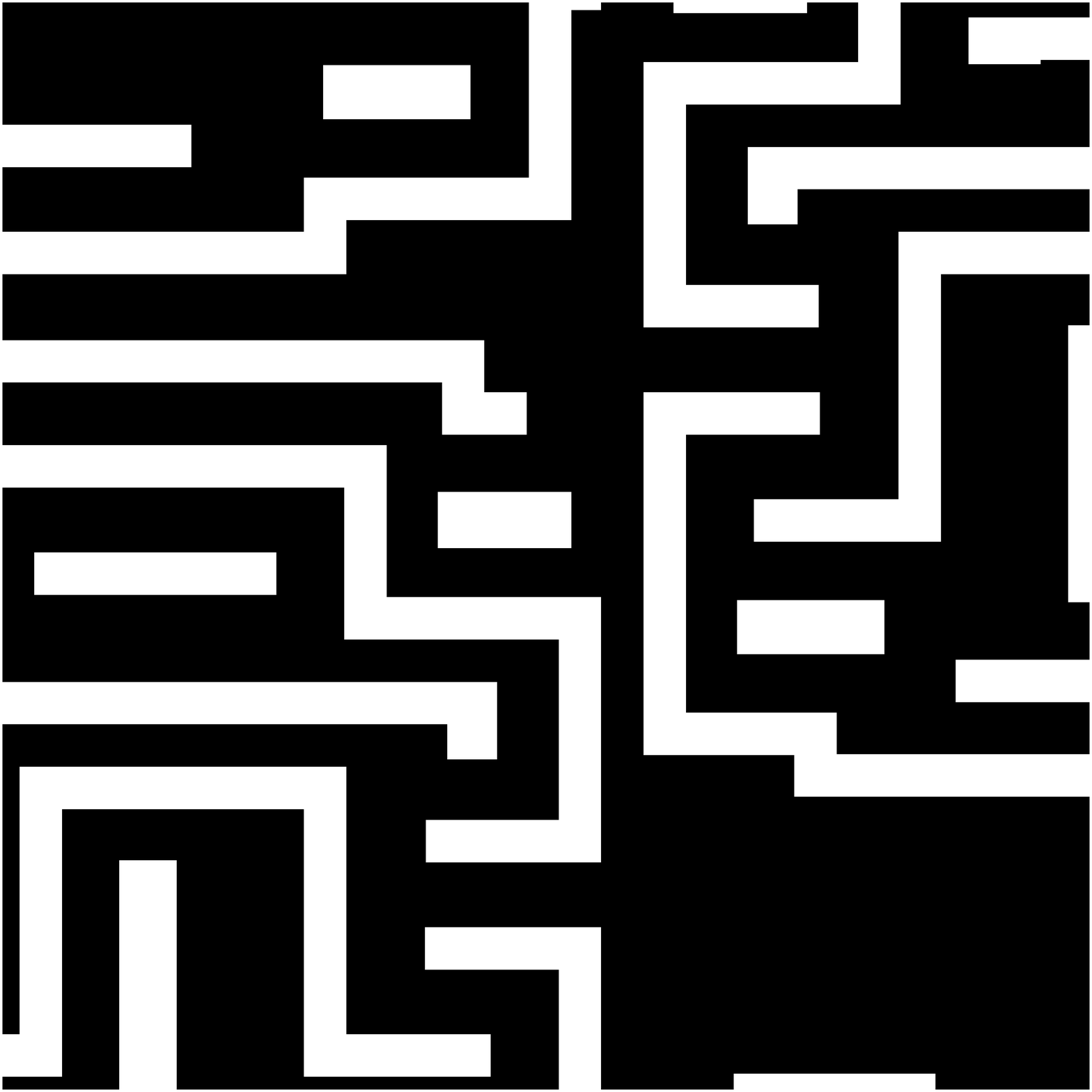}}
    \end{minipage}
    \hfill
    \begin{minipage}{0.25\linewidth}
      \centering
      \centerline{\includegraphics[width=\linewidth]{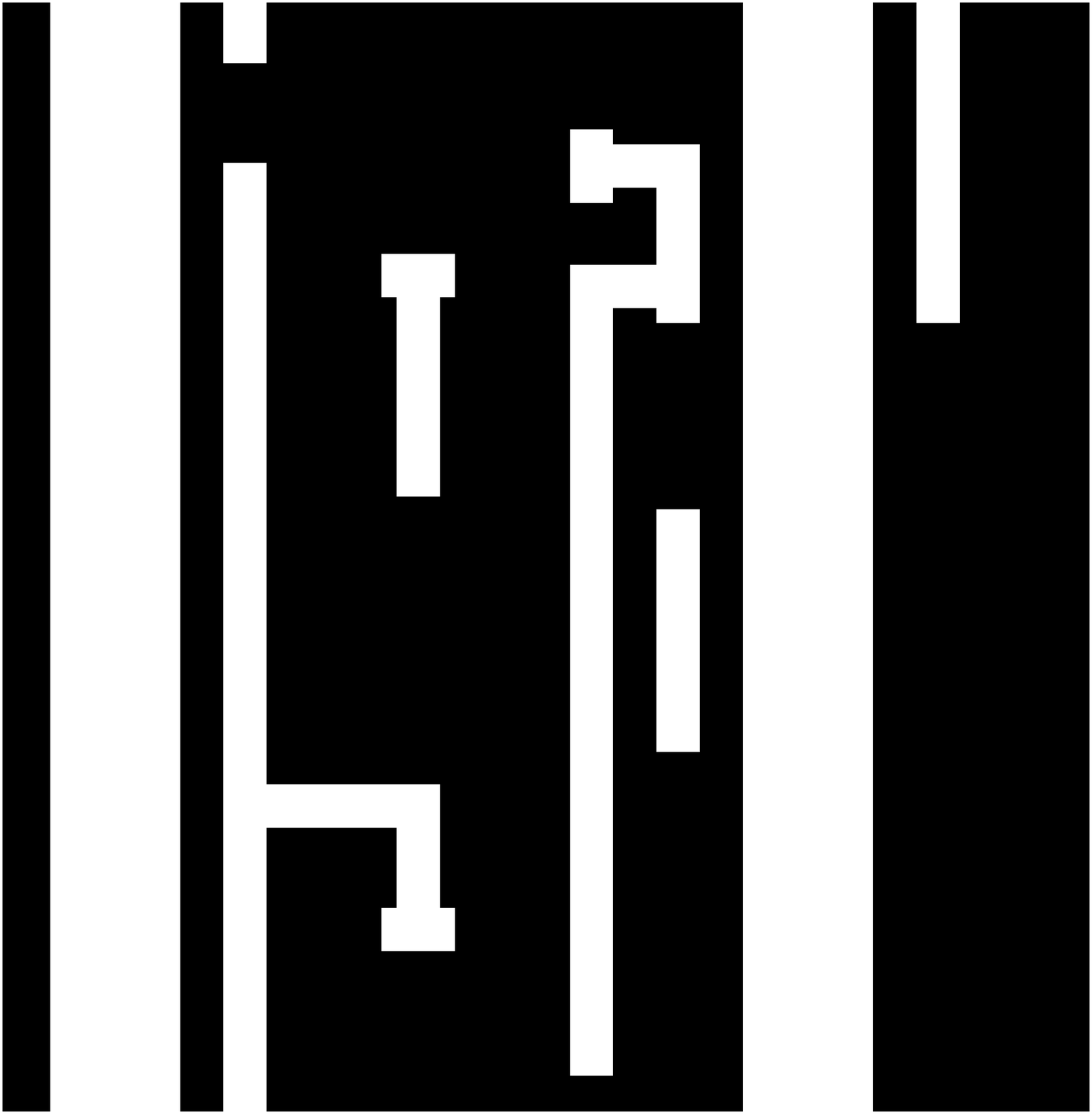}}
    \end{minipage}
    \hfill
    \begin{minipage}{0.25\linewidth}
      \centering
      \centerline{\includegraphics[width=\linewidth]{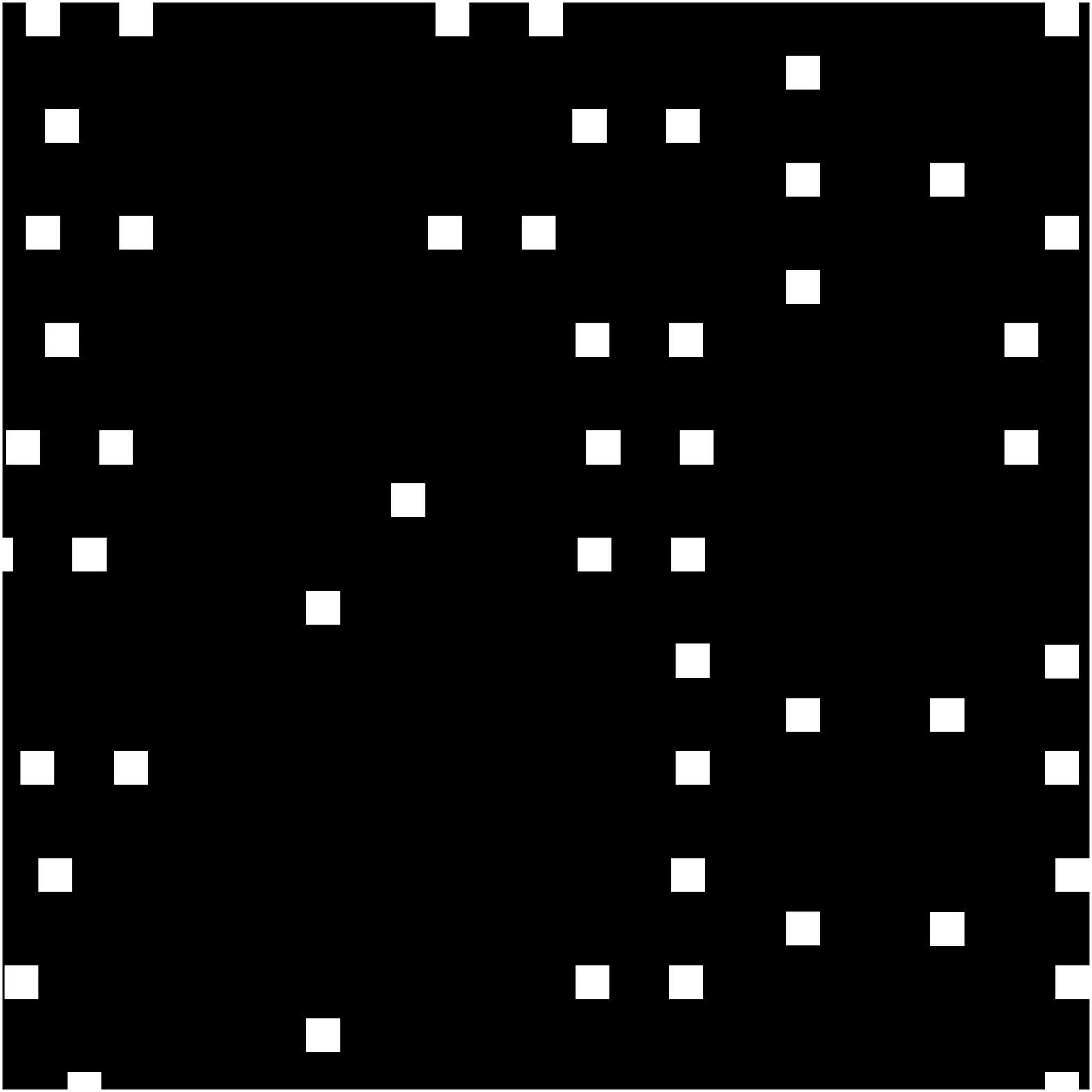}}
    \end{minipage}
  }
  \caption{From left to right: Examples of 1024nm$\times$1024nm tiles from Metal 1,
           Metal 2 and Contact Array layers, respectively. Note that they exclusively
           contain rectilinear polygons.}
  \flabel{layout}
\end{figure}

\subsubsection{Rectilinear Polygons}

We now give a mathematical definition of the polygons in VLSI layouts.  They
are rectilinear (only right angles), simple (edges do not intersect, no holes),
lattice (vertices are on the integer lattice) polygons. An example of such a
polygon is shown in \ffref{rectpoly}.

A standard layout description consists of polygons defined by the set of the
ordered coordinates of their $K$ vertices
\begin{equation}
\left\{(x_0,y_0), \ldots, (x_{K-1},y_{K-1})\right\}, \qquad (x_i, y_i) \in \mathbb{Z}^2.
\elabel{polydef_R}
\end{equation}
This set separates the plane in two: inside and outside of the polygon.  A
sequence order is also needed, which we arbitrarily choose to be clockwise. A
disjoint partition of the plane is achieved when we exclude edges if the
interior of the polygon is on their left or bottom and include them
if the interior is on the right or top (see \ffref{rectpoly}).

\begin{figure}[!t]
    \begin{center}
      \includegraphics[width=\linewidth]{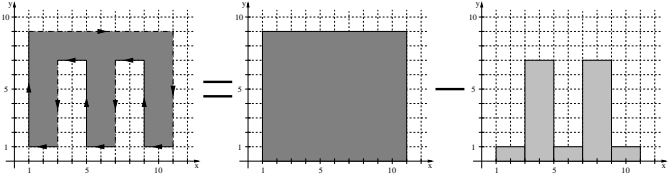}
    \end{center}
    \caption{On the left, a rectilinear simple lattice polygon. The interior of the polygon is
    shaded. The full-lined edges are included, while the dashed
    edges are not. Arrows indicate vertex ordering.
    On the right, illustration of the construction of the left polygon
            from disjoint rectangles. The minus here stands for
            the set difference operator.}
    \flabel{rectpoly}
\end{figure}

Rectangles are rectilinear polygons with four vertices, and as they are simpler
to handle, we treat them as a special case. A rectangle $\mathcal{R}$ can be
defined by its lower left and upper right vertices
\begin{equation}
\mathcal{R}_{(x_1,y_1)}^{(x_2,y_2)} = \{ (x,y)\, |\, x_1 \leq x < x_2, y_1 \leq y < y_2\}.
\nonumber
\end{equation}
The following result, illustrated in \ffref{rectpoly}, makes it easy to take
inner products over rectilinear polygons.
\begin{proposition}
Any rectilinear polygon $\Pol$ with $K$ vertices as in \eref{polydef_R} can be
expressed by $K/2$ disjoint rectangles, each with a side lying on the $x$-axis
\begin{equation}
\mathcal{P} = \underset{\{i|x_{i+1} > x_i\}}{\bigcup}\mathcal{R}_{(x_i,0)}^{(x_{i+1},y_i)}
      \Bigg\backslash \underset{\{i|x_{i+1} < x_i\}}{\bigcup}\mathcal{R}_{(x_{i+1},0)}^{(x_i,y_i)}
      \elabel{polyrectX}
\end{equation}
where the indices are taken modulo $K$ and $\backslash$ is the set difference
operator.
\label{rect_union}
\end{proposition}
An equivalent result exists for the case where a side is lying on the $y$-axis.

So far polygons have been defined as subsets of $\mathbb{R}^2$, and are not yet
directly transformable as 2D signals on $\mathbb{R}^2$. This is achieved by
transforming their indicator function
\begin{equation}
f_\mathcal{P}(x,y) = \indic{\mathcal{P}}(x,y)
\nonumber
\end{equation}
where $\mathcal{P} \subset \mathbb{R}^2$ is the polygon and
$\indic{\mathcal{S}}(x,y) = 1$, if $(x,y)\in\mathcal{S}$ and $0$ otherwise, is
the indicator function of a set $\mathcal{S}$.

\subsection{Continuous Haar Transform}
\slabel{bak_haar}

The 2D Haar basis over $\fat{T} = [0,N_x) \times [0,N_y)$ is
\begin{equation}
\left\{ \varphi_{0,0,0},\, \psi^{(hg)}_{j,k_x,k_y},\, \psi^{(gh)}_{j,k_x,k_y},\, \psi^{(hh)}_{j,k_x,k_y} \right\}
\end{equation}
where $j\in\N$ and $k_x,k_y\in\{0,\ldots,2^j-1\}$. In practice, $j$ is limited
to some maximum level of decomposition $J$. The scaling function is
\begin{equation}
\varphi_{j,k_x,k_y}(x,y) = \frac{2^j}{\sqrt{N_x N_y}}
	\varphi\left( \frac{2^j}{N_x}x - k_x \right)
	\varphi\left( \frac{2^j}{N_y}y - k_y \right)
\nonumber
\end{equation}
where $\varphi(t)=1$ if $0\leq t < 1$ and 0 otherwise. The three other basis
functions can be defined using a recursive relationship. For example
\begin{equation}
\psi^{(hg)}_{j,k_x,k_y}(x,y) = \sum\limits_n \sum\limits_m h_n g_m \varphi_{j+1,2k_x+n,2k_y+m}(x,y)
\nonumber
\end{equation}
where $g_n = [2^{-1/2} 2^{-1/2}]$ and $h_n = [2^{-1/2} -2^{-1/2}]$ are the Haar
filters.  By replacing $h_ng_m$ in the sum by $g_ng_m$, $g_nh_m$ and $h_nh_m$
we obtain $\varphi_{j,k_x,k_y}$, $\psi^{(gh)}_{j,k_x,k_y}$ and
$\psi^{(hh)}_{j,k_x,k_y}$, respectively. The dyadic CHT of a function $f$ is
given by its inner product with the basis functions. The discrete counterpart
of the CHT is the discrete Haar transform (DHT). For a more thorough
introduction to the CHT and DHT, see \cite{vetterli_world_2009}. The CHT and
DHT coefficients are identical for 2D rectilinear polygonal patterns. Both can
be computed using the fast orthogonal wavelet transform (FWT)
\cite{mallat_wavelet_2008}. This algorithm has a Cooley-Tukey butterfly
structure \cite{ahmed_orthogonal_1975}, where only the inner products with the
scaling function at the lowest level need be computed.  The full flow diagram
for a length-8 1D FWT is shown in light gray in \ffref{sigflow}.
\ffref{2Dbutterfly} shows one butterfly of the 2D transform.

\begin{figure}[tb]
  \centering
  \centerline{\includegraphics[width=0.8\linewidth]{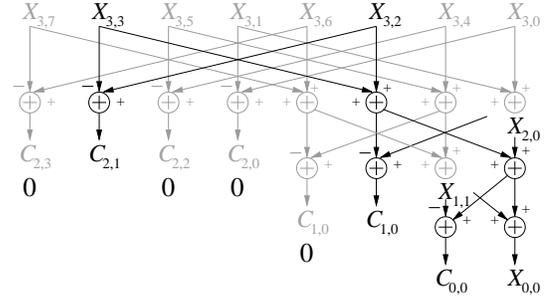}}
  \caption{A pruned signal flow of the 1D Haar FWT. The full flow-diagram is
           shown in light gray. The transformed signal is $f(t) = u(t-3)$, 
           defined on $[0,8)$, where $u(t)$ is the Heaviside function.
           $X_{j,k} = \ip{f}{\varphi^{(8)}_{j,k}}$ and $C_{j,k} = \ip{f}{\psi^{(8)}_{j,k}}$.
	   $\varphi^{(8)}_{j,k}$ and $\psi^{(8)}_{j,k}$ are the 1D Haar basis functions on $[0,8)$.
           For simplicity, the scaling of the transform coefficients has been omitted.}
  \flabel{sigflow}
\end{figure}

\begin{figure}[tb]
  \begin{center}
    \includegraphics[width=.45\linewidth]{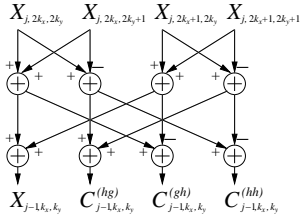}
  \end{center}
  \caption{One butterfly of the traditional 2D Haar FWT where $X_{j,k_x,k_y} =
           \ip{f}{\varphi_{j,k_x,k_y}}$ and $C^{(a,b)}_{j,k_x,k_y}=\ip{f}{\psi^{(ab)}_{j,k_x,k_y}}$, $j\in\N$,
           $k_x,k_y\in\{0,\ldots,2^j-1\}$, and $a,b\in\{g,h\}$. Scaling of
           the transform coefficients is omitted for simplicity.}
  \flabel{2Dbutterfly}
\end{figure}

\subsection{Continuous Fourier Series}

The 2D Fourier basis over $\fat{T} = [0,N_x) \times [0,N_y)$ is
\begin{equation}
\left\{ (N_x N_y)^{-1/2} e^{j (w_x k x + w_y l y)} \right\}_{(k,l)\in\mathbb{Z}^2},
\elabel{fourier_basis}
\end{equation}
where $w_x = \frac{2\pi}{N_x}$ and $w_y = \frac{2\pi}{N_y}$. The Fourier basis
assumes that the function $f$ under transformation is periodic with period
$N_x$ along the $x$-axis and period $N_y$ along the $y$-axis. The CFS
coefficients $\hat{F}_{k,l}$ are then given by the inner product between $f$
and the Fourier basis functions.

The main difference with the DFT is that the functions are continuous in the
spatial domain, and thus not periodic in the frequency domain. This means that
for a perfect reconstruction of the image, an infinite number of coefficients
is needed. On the other hand, the use of the DFT requires sampling, which in
turn introduces aliasing, due to the infinite bandwidth of rectilinear
polygons. In contrast, the CFS yields the true spectrum of the continuous
image.

\section{Continuous Transforms of Rectilinear Polygons}
\slabel{algo}

In this section, we derive the algorithms to compute the continuous Haar
transform and Fourier series coefficients of rectilinear polygons. Their
theoretical computational complexity is evaluated and compared with that of
the equivalent discrete transforms.

The main reason continuous transforms can be faster than their discrete
counterparts for rectilinear polygons is that the continuous inner product of a
basis function is an explicit function of the vertices of the polygon, and the
vertex description \eref{polydef_R} is very sparse compared to the size of the
image. Moreover, no memory is needed to form or store a discrete image.  On the
other hand, the sampling of the polygons to create a discrete image can itself
be considered a projection on a Dirac basis. Sampling followed by a discrete
transform, as illustrated in \ffref{transflow}, effectively is two transforms,
whereas a continuous transform can completely omit the sampling operation.
Moreover, as discussed in the introduction, sampling introduces aliasing since
the assumption that the signal is bandlimited does not hold for rectilinear
polygons.

\begin{figure}[tb]
  \centering
  \centerline{\includegraphics[width=\linewidth]{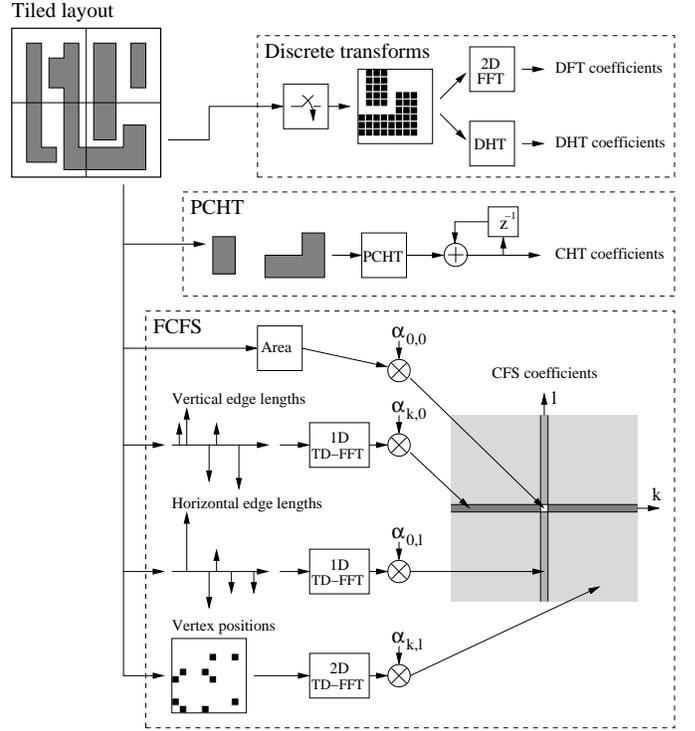}}
  \caption{Flow diagram of all the transforms. From top to bottom: fast Fourier 
           transform (FFT), discrete Haar transform (DHT), pruned continuous Haar
           transform (PCHT) and fast continuous Fourier series (FCFS).}
  \flabel{transflow}
\end{figure}

\subsection{Inner Product over Rectilinear Polygons}

In practice, layouts are divided into smaller disjoint or overlapping
rectangular tiles before applying a transform to each individual tile.
Therefore, we consider continuous transforms over a rectangular subset $\fat{T}
= [0,N_x)\times[0,N_y) \subset \mathbb{R}^2$ that we call a tile. This poses no
restrictions as the size can be increased sufficiently to cover the entire
layout. In the case of VLSI layouts, $N_x,N_y$ are always chosen to be positive
integers. An orthogonal basis of functions $\phi_{k,l}$ over $\fat{T}$ can be
written as $\left\{\phi_{k,l} : \fat{T} \longrightarrow \mathbb{C}
\right\}_{k=0,l=0}^{\infty,\infty}$, and the transform coefficients of a
function $f\in L^2(\fat{T})$ are simply the inner products with the basis
functions
\begin{equation}
\ip{f}{\phi_{k,l}} = \iint\limits_{\fat{T}} f(x,y)\phi_{k,l}^*(x,y)\,dx\,dy.
\elabel{inner_prod}
\end{equation}

As all polygons in a layout are disjoint, the image $f_T$ of a tile
containing polygons $\mathcal{P}_0$, \dots , $\mathcal{P}_{M-1} \subseteq
\fat{T}$ is the sum of the indicator functions of the polygons
\begin{equation}
f_T(x,y) = \sum\limits_{m=0}^{M-1} \indic{\mathcal{P}_m}(x,y).
\elabel{signal_model}
\end{equation}
Now that we have a signal model, we provide two results that will allow us to
derive the inner product of $f_T$ with a given basis function.
\begin{proposition}
If $f_T(x,y)$ is given by \eref{signal_model} and $\phi_{k,l}(x,y)$ is a basis
function, their inner product according to \eref{inner_prod} is
\begin{equation}
\ip{f_T}{\phi_{k,l}} = \sum\limits_{m=0}^{M-1} \sum\limits_{i=0}^{K_m/2-1} \int\limits_0^{y_{m,2i}} \int\limits_{x_{m,2i}}^{x_{m,2i+1}} \phi_{k,l}^*(x,y)\,dx\,dy
\nonumber
\end{equation}
where $(x_{m,i},y_{m,i})$ is the $i$\textsuperscript{th} vertex, out of the
$K_m$, of the $m$\textsuperscript{th} polygon and $i$ is taken modulo $K_m$.
Here it is assumed without loss of generality that for all $m\ x_{m,0} \neq
x_{m,1}$.
\label{poly_int}
\end{proposition}
\begin{IEEEproof}
By the linearity of the inner product and since the polygons are disjoint, we
have
\begin{equation}
\ip{f_T}{\phi_{k,l}} = \sum\limits_{m=0}^{M-1} \ip{f_{\Pol_m}}{\phi_{k,l}} = \sum\limits_{m=0}^{M-1} \iint\limits_{\mathcal{P}_m} \phi_{k,l}^*(x,y)\,dx\,dy.
\elabel{poly_in_prod}
\end{equation}
Then, using Proposition \ref{rect_union}, we split the integral over
$\Pol_m$ into $K_m/2$ over rectangles. If $x_{m,i} < x_{m,i+1}$, the integral is
positive and is added to the final result. If $x_{m,i} > x_{m,i+1}$, the
integral is negative and is subtracted from the final result. If $x_{m,i} =
x_{m,i+1}$, the $i$\textsuperscript{th} term of the sum is zero.
\end{IEEEproof}

We now give the closed-form formula for $\ip{f_{\Pol}}{\phi_{k,l}}$ when
$\phi_{k,l}$ is separable. When there is more than one polygon in the tile, we
compute the continuous Haar and Fourier transforms of $f_T$ by simply applying
\eref{poly_in_prod}. The following lemma is derived using
Proposition~\ref{poly_int}, the separability of $\phi_{k,l}$ and the fact that
$\sum_{i=0}^{K/2-1} \left(\Omega_k^*(x_{2i+1}) - \Omega_k^*(x_{2i}) \right) =
0$ for rectilinear polygons since the index $i$ is cyclic and vertical edges
cancel. The function $\Omega$ is a primitive of $\phi$.
\begin{lemma}
Consider a rectilinear polygon $\Pol$ with $K$ vertices, and a separable basis
function, namely $\phi_{k,l}(x,y) = \phi_k(x)\phi_l(y)$. Then
\begin{equation}
\ip{f_{\Pol}}{\phi_{k,l}} = \sum\limits_{i=0}^{K/2-1} \Omega_l^{*}(y_{2i}) \left(\Omega_k^{*}(x_{2i+1}) - \Omega_k^{*}(x_{2i}) \right),
\elabel{polyip}
\end{equation}
where indices are taken modulo $K$.
\label{polyip}
\end{lemma}

\subsection{Pruned Continuous Haar Transform (PCHT)}

We first describe the PCHT algorithm for rectilinear polygons, first
introduced in \cite{scheibler_pruned_2010}. We then advance this previous work
with the derivation of the complexity of PCHT.

\subsubsection{Algorithm Derivation}
\slabel{haar_algo}

Consider the FWT as described in \sref{bak_haar} and whose full 1D flow diagram
is shown in light gray in \ffref{sigflow}. First, note that the linear
complexity of the FWT and \eref{poly_in_prod} mean that we can use the FWT on
individual polygons and sum up the transform coefficients to obtain the
transform of a tile, as illustrated in \ffref{transflow}. Thus, from now on, we
consider only the transform of a single polygon. Second, we use computational
geometry techniques to compute the inner product. The continuous inner product
between the indicator of a polygon and the support of the scaling function, as
given by Lemma~\ref{polyip}, is the area of the geometrical intersection of the
polygon and the scaling function multiplied by $2^j/\sqrt{N_x N_y}$. A method
to compute the intersection area for rectilinear polygons is described in
\algref{intarea}.

The Haar transform acts as a discontinuity detector, and all transform
coefficients will be zero except when basis functions intersect the boundary of
the polygon. Therefore, the basis functions completely inside or outside the
polygon can be ignored. In addition, all coefficients positioned below such a
basis function in the transform tree are also zero. \ffref{pru_ip} shows, in
gray, the support of the basis functions of a given scale that yield non-zero
inner products. As with the original FWT, PCHT can be written as a
divide-and-conquer algorithm: Divide the tile into four rectangular parts
recursively until the part under consideration is completely inside or outside
the polygon. Pseudocode for PCHT can be found in \cite{scheibler_pruned_2010}.
An example of the pruned transform flow-diagram for the 1D case is shown in
\ffref{sigflow} in black.

\begin{figure}[t]
  \begin{center}
    \includegraphics[width=0.35\linewidth]{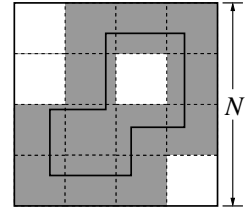}
  \end{center}
    \caption{The Haar basis functions yield zero inner product except when they
            intersect the edge of a polygon. The big thick square is the tile with $N_x =
            N_y = N$. The dashed squares are the supports of the basis functions at a given
            scale. The thick polygonal border marks the contour of the polygon. Basis
            functions yielding non-zero inner product are shown in gray.}
  \flabel{pru_ip}
\end{figure}

\subsubsection{Complexity}
\slabel{haar_comp}

We will now estimate the complexity of PCHT. As it is highly dependent on the
geometry of the polygon, we make the worst-case assumption and describe the
complexity as a function of a few general properties of the polygon, such as
its number of vertices $K$ and its perimeter $P$. The complexity is also a
function of the computational complexity $\Lambda_{ia}(K)$ of the intersection
area algorithm. The exact value depends on the assumptions made regarding the
type of polygons, but is $O(K)$. In particular, for rectilinear polygons, we
have
\begin{equation}
\Lambda_{ia}(K) = 
  \begin{cases}
    11 & \text{if $K=4$} \\
    6K-1 & \text{if $K>4$}
  \end{cases},
\elabel{int_comp}
\end{equation}
additions, multiplications and comparisons.

We begin by estimating the number of non-terminating recursive calls. A call is
non-terminating if the support of the waveform intersects the boundary of the
polygon. At a given scale $j$, there are roughly $\ceil{2^jP/N}$ such basis
functions. We assume here that $N_x=N_y=N$. Each of these calls makes four
calls to itself and thus to the intersection area routine and also uses three
comparisons. The intersection area routine uses in turn eleven additions and
three multiplications per non-terminating call. Finally, a total of $M-1$
additions are needed to sum up the scaling coefficients of the polygons, and a
single multiplication is needed for the final scaling factor. We thus formulate
our estimate of the complexity of PCHT as
\begin{equation}
\Lambda_{PCHT} \approx \left(\sum\limits_{m=0}^{M-1} \sum\limits_{j=0}^{J-1} 
          \ceil{\frac{2^jP_m}{N}} \left( \Lambda_{ia}(K_m)+26 \right)\right) + M
\elabel{haar_comp}
\end{equation}
operations (additions, multiplications and comparisons), where $P_m$ and $K_m$
are the perimeter and number of vertices of the $m$\textsuperscript{th}
polygon, respectively, $M$ is the number of polygons in a given tile and $J$
the maximum level of decomposition. For a DHT with $N_x=N_y=N=2^J$, the total
number of operations is $\Lambda_{DHT} = \frac{8}{3}(N^2-1)$. The relation
between the complexity and the actual runtime is discussed in
\sref{algo_perf_eval}.

\begin{algorithm}[t]
  \caption{IntersectionArea$\left(\Pol,\fat{T}_{j,k_x,k_y}\right)$} \alabel{intarea}
    \begin{algorithmic}[1] 
    \REQUIRE A rectilinear polygon $\Pol$. The support $\fat{T}_{j,k_x,k_y}$ of the basis function $\varphi_{j,k_x,k_y}$.
    \ENSURE $I$ is the intersection area of $\Pol$ and $\fat{T}_{j,k_x,k_y}$.

    \STATE $I \gets 0$
    \STATE $a_1 \gets k_xN_x/2^j$, $b_1 \gets k_yN_y/2^j$
    \STATE $a_2 \gets (k_x+1)N_x/2^j$, $b_2 \gets (k_y+1)N_y/2^j$

    \FOR{every horizontal edge $(x_i,y_i) \rightarrow (x_{i+1},y_i)$ of $\Pol$}
      \STATE $u \gets \min(x_i,x_{i+1})$,  $v \gets \max(x_i,x_{i+1})$
      \IF{$a_2\leq u$ or $v\leq a_1$ or $y_i\leq b_1$}
        \STATE continue
      \ENDIF
      \STATE $s \gets \operatorname{sign}(x_{i+1}-x_i)$
      \STATE $I \gets I + s (\min(v,a_2) - \max(u,a_1)) (\min(y_i,b_2) - b_1)$
    \ENDFOR

    \end{algorithmic}
\end{algorithm}

\subsection{Fast Continuous Fourier Series (FCFS)}

\begin{table*}[tb]
  \caption{Complexity of the steps of the FCFS
           algorithm. $K=\sum_{m=0}^{M-1}K_m$ is the sum of the vertices of the $M$
           polygons in the $N_x \times N_y$ tile. The parameters $P_x^{(I)}$, $P_x^{(II)}$,
           dividers of $N_x$, and $P_y^{(I)}$, $P_y^{(II)}$, dividers of $N_y$, are chosen
           to minimize the overall complexity.}
  \tlabel{fourier_comp}
\begin{center}
  \begin{tabular}{@{}l@{~~~}l@{~~~}l@{}}
    \toprule
    \textbf{Step} & \textbf{Number of Operations} & \textbf{Operation performed} \\
    \midrule
    1 & $\frac{3}{2}K - M$ & Sum of polygon areas \\
    2 & $\big(\frac{5}{2} + \frac{3}{2}Q_x^{(I)}\big) \times K + \big(\frac{1}{2}Q_x^{(I)}+1\big) \times C_{FFT1D}\big(P_x^{(I)}\big)^\dagger$ & 1D $\frac{K}{2}$-input pruned FFT \\
    3 & $\big(\frac{5}{2} + \frac{3}{2}Q_y^{(I)}\big) \times K + \big(\frac{1}{2}Q_y^{(I)}+1\big) \times C_{FFT1D}\big(P_y^{(I)}\big)^\dagger$ & 1D $\frac{K}{2}$-input pruned FFT \\
    4 & $\big(-\frac{17}{2} + \frac{19}{2}Q_x^{(II)} + \frac{5}{2}Q_x^{(II)}Q_y^{(II)}\big) \times K + Q_x^{(II)} \big(\frac{1}{2}Q_y^{(II)}+1\big) \times C_{FFT2D}\big(P_x^{(II)}, P_y^{(II)}\big)^\ddagger$ & 2D $K$-input pruned FFT \\
    \midrule
    \textbf{Total} & \multicolumn{2}{@{}l@{}}{$\big( -2+\frac{3}{2}Q_x^{(I)}+\frac{3}{2}Q_y^{(I)}+\frac{19}{2}Q_x^{(II)}+\frac{5}{2}Q_x^{(II)}Q_y^{(II)}\big)K-M$} \\
                   & \multicolumn{2}{@{}c@{}}{$+\, \big(\frac{1}{2}Q_x^{(I)}+1\big)\times C_{FFT1D}\big(P_x^{(I)}\big) + \big(\frac{1}{2}Q_y^{(I)}+1\big) \times C_{FFT1D}\big(P_y^{(I)}\big)$} \\
                   & \multicolumn{2}{@{}r@{}}{$+\, Q_x^{(II)}\big(\frac{1}{2}Q_y^{(II)}+1\big)\times C_{FFT2D}\big(P_x^{(II)}, P_y^{(II)}\big)$ Operations} \\
    \bottomrule
    \multicolumn{3}{@{}r@{}}{ {\tiny \textsuperscript{$\dagger$}1D Split-Radix: $C_{FFT1D}(N)=4N\log_2N-6N+8$ \hspace{3pt}
                                     \textsuperscript{$\ddagger$}2D Split-Radix: $C_{FFT2D}(N_x,N_y)=4N_xN_y\log_2N_xN_y-12N_xN_y+8N_x+8N_y$ } }
  \end{tabular}
\end{center}
\end{table*}

\begin{table*}[tb]
  \caption{Summary of the computational complexity of the continuous and
          discrete transforms of a single rectilinear polygon in terms of
          the total number of additions and multiplications.}
  \tlabel{sum_comp}
\begin{center}
  \begin{tabular}{@{}lcclc@{}}
    \toprule
    \multicolumn{4}{@{}l@{}}{\textbf{Complexity of the Transforms:}} \\
    \midrule
    \textbf{PCHT} & \multicolumn{2}{@{}c@{}}{$\approx {\left(\sum\limits_{i=0}^{M-1} \sum\limits_{j=0}^{J-1} 
          \ceil{\frac{2^jP_i}{N}} \left( \Lambda_{ia}(K_i)+26 \right)\right) + M}\;^\dagger$} & \textbf{DHT} & $8\frac{N^2 - 1}{3}$ \\
    & {\tiny \textbf{Goertzel}} & {\tiny \textbf{TD-FFT}} & \\
    \cmidrule{2-3}
    \textbf{FCFS} & $\big(\big(N_x+\frac{1}{4}\big)\big(N_y+\frac{5}{4}\big)+\frac{19}{16}\big)K-M$ & See \tref{fourier_comp}   
                                                                                              & \textbf{FFT} & $\frac{1}{2}\big(4N_xN_y\log_2 N_xN_y-12N_xN_y+8N_x+8N_y\big)$ $^\ddagger$ \\
    \bottomrule
    \multicolumn{5}{@{}r@{}}{{\tiny  \textsuperscript{$\dagger$}$\Lambda_{ia}(K_i)$ given in \eref{int_comp} \hspace{3pt}
                                     \textsuperscript{$\ddagger$}Row-column real split-radix FFT complexity \cite{duhamel_split_1984}}}
  \end{tabular}
\end{center}
\end{table*}

We now develop the FCFS algorithm for computing the CFS coefficients of
rectilinear polygons. The computational complexity of FCFS will also be
evaluated and compared with that of the FFT.

\subsubsection{Algorithm Derivation}
\slabel{fourier_algo}

We begin by a closed-form formula for the CFS coefficients of rectilinear
polygons. Applying Lemma~\ref{polyip} to the Fourier basis
\eref{fourier_basis} results in the following proposition.
\begin{proposition}
Given a polygon $\Pol$ with $K$ vertices, its CFS $\hat{F}_{k,l}$, $k,l\in\Z^2$, is given by
\begin{equation}
\hat{F}_{0,0} = \alpha_{0,0} \sum\limits_{i=0}^{K/2-1} y_{2i} (x_{2i+1} - x_{2i}),
\elabel{Fdc}
\end{equation}
\begin{equation}
\hat{F}_{k,0} = \alpha_{k,0} \sum\limits_{i=0}^{K/2-1} (y_{2i-1} - y_{2i}) e^{-j w_x k x_{2i}},
\elabel{Fk0}
\end{equation}
\begin{equation}
\hat{F}_{0,l} = \alpha_{0,l} \sum\limits_{i=0}^{K/2-1} (x_{2i+1} - x_{2i}) e^{-j w_y l y_{2i}},
\elabel{F0l}
\end{equation}
\begin{equation}
\hat{F}_{k,l} = \alpha_{k,l} \sum\limits_{i=0}^{K/2-1} e^{-j w_y l y_{2i}}\left(e^{-j w_x k x_{2i+1}} - e^{-j w_x k x_{2i}}\right),
\elabel{Fkl}
\end{equation}
where the scaling factor $\alpha_{k,l}$ is defined as follows
\begin{equation}
\alpha_{k,l} = \begin{cases}
  \frac{1}{\sqrt{N_x N_y}} & \text{if $k=l=0$} \\
  \frac{j}{2\pi k} \sqrt{\frac{N_x}{N_y}} & \text{if $k\neq0$ and $l=0$} \\
  \frac{j}{2\pi l} \sqrt{\frac{N_y}{N_x}} & \text{if $k=0$ and $l\neq0$} \\
  -\frac{\sqrt{N_x N_y}}{4 \pi^2 k l} & \text{if $k\neq0$ and $l\neq0$}
\end{cases}.
\nonumber
\end{equation}
\label{CFT}
\end{proposition}

It turns out that, except for $\hat{F}_{0,0}$, all coefficients can be computed
using DFTs of very sparse signals as we restrict the vertices to lie on the
integer lattice. We can decompose the algorithm into four steps, with each
step computing one of \eref{Fdc} to \eref{Fkl}. For simplicity, we consider
only the transform of a single polygon here. 
\begin{trivlist}
  \item[\textit{\textbf{Step 1:}}] Directly compute \eref{Fdc}. It is
    the scaled area of $\Pol$. $\hat{F}_{0,0} = \alpha_{0,0}
    \operatorname{Area}\left(\Pol\right).$ This corresponds to line 11 and 18 in the \algref{Fourier}.

  \item[\textit{\textbf{Step 2:}}] Compute \eref{Fk0} as a 1D DFT: $\hat{F}_{k,0} =
    \alpha_{k,0} \operatorname{DFT}_{k^\prime}\left\{ \tilde{f}_x \right\}$, where
    $k^\prime \equiv k \bmod N_x$ and
    \begin{equation}
      \tilde{f}_x[n] = \sum\limits_{i\in \mathcal{X}_n} (y_{i-1} - y_i),\quad n=0,\ldots,N_x-1, 
      \nonumber
    \end{equation}
    where $\mathcal{X}_n = \{i\,|\,x_i\equiv n \bmod N_x\}$. The values
    $(y_{i-1} - y_i)$ are the lengths of the vertical edges. This corresponds to line 5 and 16 in the \algref{Fourier}.

  \item[\textit{\textbf{Step 3:}}] Compute \eref{F0l} as a 1D DFT:
    $\hat{F}_{0,l} = \alpha_{0,l} \operatorname{DFT}_{l^\prime}\left\{ \tilde{f}_y
    \right\}$, where $l^\prime \equiv l \bmod N_y$ and
    \begin{equation}
      \tilde{f}_y[n] = \sum\limits_{i\in \mathcal{Y}_n} (x_{i+1} - x_i),\quad n=0,\ldots,N_y-1,
      \nonumber
    \end{equation}
    where $\mathcal{Y}_n = \{i\,|\,y_i\equiv n\bmod N_y\}$. The values
    $(x_{i+1} - x_i)$ are the lengths of the horizontal edges. This corresponds to line 8 and 17 in the \algref{Fourier}.

  \item[\textit{\textbf{Step 4:}}] Compute \eref{Fkl} as a 2D DFT:
    $\hat{F}_{k,l} = \alpha_{k,l} \operatorname{DFT}_{k^\prime,l^\prime} \left\{
    \tilde{f}_{xy} \right\}$, where $k^\prime \equiv k \bmod N_x$,
    $l^\prime \equiv l \bmod N_y$ and
    \begin{equation}
      \tilde{f}_{x,y}[m,n] = \sum\limits_{i=0}^{K/2-1} 
        \left( \indic{\mathcal{Z}_{m,n}}(x_{2i+1}, y_{2i}) 
        - \indic{\mathcal{Z}_{m,n}}(x_{2i},y_{2i}) \right)
      \nonumber
    \end{equation}
    where $\mathcal{Z}_{m,n} = \big\{ (x,y)\, |\, (x,y)\in\Z^2,\, x\equiv m
    \bmod N_x,\, y\equiv n \bmod N_y \big\}$.
    This is a sparse $N_x\times N_y$ image with $1$'s and $-1$'s placed at the
    vertices. This corresponds to line 8, 9 and 15 in the \algref{Fourier}.
\end{trivlist}
For multiple polygons, we can use \eref{poly_in_prod} for Step 1, and the
linearity of the DFT is used to include all the polygons in $\tilde{f}_x$,
$\tilde{f}_y$ and $\tilde{f}_{xy}$ for the three other steps. The four steps
are illustrated in \ffref{transflow}. \algref{Fourier} gives the pseudocode of
FCFS. In practice, it is often possible to combine the creation of
$\tilde{f}_x$, $\tilde{f}_y$ and $\tilde{f}_{xy}$ with the computation of the
DFTs, in which case no full-length arrays need to be stored.

Having reduced the problem of computing the CFS to a few real DFTs of sparse
signals, we can exploit the extensive collection of available DFT algorithms.
The FFT algorithm \cite{cooley_algorithm_1965} is not the most efficient in our
case, as we could also use a single FFT on a sampled version of $f_T$, albeit
with some loss of precision. If we are only interested in a few CFS
coefficients, we can apply a Goertzel-like algorithm
\cite{goertzel_algorithm_1958} to the direct computation of the CFS
coefficients using Proposition~\ref{CFT}. The pruned FFT is the most appealing
approach for the computation of a large number of CFS coefficients, and in
particular the transform decomposition FFT (TD-FFT) as it is the fastest of all
existing pruned FFT algorithms~\cite{sorensen_efficient_1993},
\cite{medina-melendrez_input_2009}. The complexity analysis considers both
TD-FFT and Goertzel, while the implementation is focused solely on TD-FFT.

\begin{algorithm}[tb]
  \caption{FCFS$(\fat{\Pol},N_x,N_y,F)$} \alabel{Fourier}
    \begin{algorithmic}[1]
    \REQUIRE $\fat{\Pol}$ contains the lists of the vertices of the $M$
             rectilinear polygons where $(x_{m,i},y_{m,i})$ is the $i$\textsuperscript{th}, out of $K_m$, vertex
             of the $m$\textsuperscript{th} polygon $\Pol_m$. For all $m\,:\,x_{m,0}\neq x_{m,1}$. The tile size is
             $N_x\times N_y$. TD-FFT-1D and -2D take as inputs the DFT length and a sparse array whose non-zero
	     entries are stored in $V$, and their locations in $S$.
    \ENSURE $F$ is an $N\times N$ matrix containing the $N\times N$ first Fourier series coefficients of the polygon $\Pol$.

    \STATE $n \gets 0$; $A \gets 0$
    \FOR{$m=0$ to $M-1$}
      \FOR{$i=0$ to $K_m-1$}
        \IF{$x_{m,i} = x_{m,i+1}$}
          \STATE $V^K[n] \gets y_{m,i} - y_{m,i+1};\ S^K[n] \gets x_{m,i}$
          \STATE $n \gets n+1$
        \ELSIF{$y_{m,i} = y_{m,i+1}$}
          \STATE $V^L[n] \gets x_{m,i+1} - x_{m,i};\ S^L[n] \gets y_{m,i}$
          \STATE $V^{KL}[2n] \gets 1;\ S^{KL}[2n] \gets (x_{m,i+1},y_{m,i})$
          \STATE $V^{KL}[2n+1] \gets -1;\ S^{KL}[2n+1] \gets (x_{m,i},y_{m,i})$
          \STATE $A \gets A + y_{m,i}(x_{m,i+1} - x_{m,i})$
        \ENDIF
      \ENDFOR
    \ENDFOR

    \STATE $F[k,l] \gets \alpha_{k,l}\operatorname{TD-FFT-2D}\left( V^{KL},S^{KL},N_x,N_y \right)$
    \STATE $F[k,0] \gets \alpha_{k,0}\operatorname{TD-FFT-1D}\left( V^K,S^K,N_x \right)$
    \STATE $F[0,l] \gets \alpha_{0,l}\operatorname{TD-FFT-1D}\left( V^L,S^L,N_y \right)$
    \STATE $F[0,0] \gets \alpha_{0,0} A$
  \end{algorithmic}
\end{algorithm}

\subsubsection{Complexity}

As in the Haar case, the computational complexity depends on polygon geometry.
In this case, only $K$, the sum of the number of vertices of the $M$ polygons
present in the tile, is important.

We compare in terms of complexity FCFS with the split-radix FFT
\cite{duhamel_split_1984} performed on the discrete image of a polygon. We
choose the split-radix FFT algorithm for it is one of the fastest and most
widely used FFT algorithms. For FCFS, we consider both Goertzel and TD-FFT
algorithms to compute the sparse DFTs.

For the complexity analysis, consider the transform of a $N_x\times N_y$ tile,
where $N_x$ and $N_y$ are composite numbers. \tref{fourier_comp} details the
complexity of each step of the FCFS algorithm described in \sref{fourier_algo}.
As observed in \cite{medina-melendrez_input_2009}, we need to choose the TD-FFT
sub-FFTs lengths $P_x^{(I)}$, $P_y^{(I)}$, $P_x^{(II)}$ and $P_y^{(II)}$ such
that the overall complexity is minimized, with the constraint that they are
dividers of $N_x$ and $N_y$, respectively. There is however no closed-form
solution to this problem. In addition, this optimization requires knowledge of
the specific FFT algorithm used for the sub-FFTs in TD-FFT, which is impossible
with modern libraries such as FFTW. Therefore, the lengths of the sub-FFTs need
to be chosen such that the runtime on a given architecture is minimized. This
can be done offline, and the optimal lengths stored in a look-up table.

For the sake of analysis, we assume that the split-radix FFT algorithm is used
for all FFTs. The complexity of the 2D $N_x\times N_y$ complex split-radix FFT
is
\begin{equation}
C_{FFT2D} = 4N_xN_y\log_2N_xN_y-12N_xN_y+8N_x+8N_y
\nonumber
\end{equation}
real operations. Real-valued data implies the need for roughly half this number
of operations. The Goertzel algorithm requires approximately $O(KN_xN_y)$ real
operations for an $N_x\times N_y$-point real DFT with $K$ non-zero inputs. Our
algorithm requires the computation of the area of the polygons (step 1), two
length-$N$ DFTs with $K/2$ inputs each (steps 2 and 3) and one 2D length-$N_x
\times N_y$ DFT with $K$ inputs (step 4). The exact complexity of FCFS is given
in \tref{fourier_comp}. \tref{sum_comp} provides a summary of the computational
complexity of all the transforms.

\begin{figure*}[tb]
  \centering
  \begin{minipage}{\linewidth}
    \footnotesize Runtime [ms] \hfill Complexity [$\times10^6$ Op] \\
    \centerline{\includegraphics[width=\linewidth]{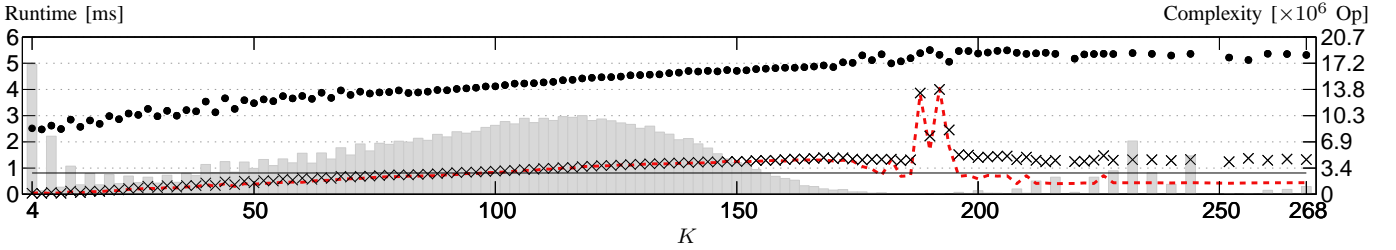}} \\
    \centerline{$K$} \\[-0.4cm]
  \end{minipage}
  \caption{Median runtime (left $y$-axis, data points) and complexity (right $y$-axis,
          lines) of the PCHT ($\times$, dashed line) and DHT ($\bullet$, solid line) of
          1024nm$\times$1024nm tiles from the M1 layer containing $K$ vertices. Note the
	  accuracy of our complexity estimate in predicting the qualitative behavior of
	  the PCHT (superposition of crosses and dashed line). Even the outliers around $K=190$ are predicted.
          Tiles with $K>200$ have a lower complexity as they contain
          only rectangles, which are less complex. However, the complexity is underestimated
          as it does not take memory transfers into account. The empirical distribution
          of the number of vertices is shown in gray.}
  \flabel{haar_comp}
\end{figure*}

\begin{figure}[tb]
  \centering
  \centerline{
    \begin{minipage}{\linewidth}
      \footnotesize Speed-up \\
      \centerline{\includegraphics[width=\linewidth]{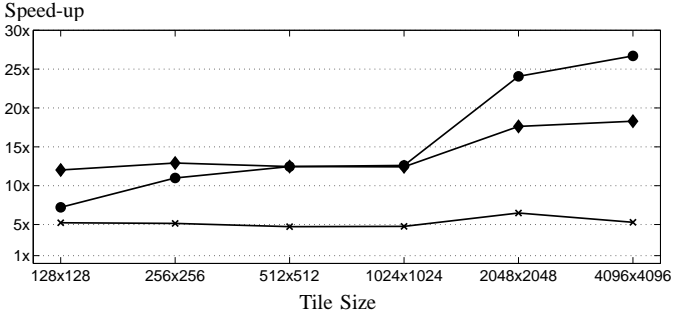}} \\
      \centerline{Tile Size} \\[-0.4cm]
    \end{minipage}
  }
  \caption{Speed-up of the average runtime provided by the PCHT on layers M1 ($\times$),
           M2 ($\blacklozenge$), CA (\mbox{\Large $\bullet$}). The higher the
           better. 1 times speed-up means same performance as the DHT.
           We see that PCHT is at least 5 times faster than the DHT. Confidence intervals have been
           omitted in this figure as in the worst case they were found to be
           within $\pm0.06$ of the value given, with 95\% confidence.}
  \flabel{haar_speedup}
\end{figure}

\section{Performance Evaluation of PCHT and FCFS}
\slabel{algo_perf_eval}

In this section, we first describe how PCHT and FCFS are implemented, and then
present results benchmarked on real VLSI layouts, comparing the performance to
that of traditional discrete transform algorithms. This performance evaluation
has two goals. The first is to validate the theoretical computational
complexity and analyze the behavior of the runtime as a function of the number
of vertices $K$ in a tile. The second goal is to measure the improvement in
runtime provided by the PCHT and FCFS over the DHT and FFT, respectively.

\subsection{Implementation and Setup of the Benchmark}

All algorithms were implemented in a computational lithography tool, and run on
a 3GHz Intel{\copyright} Xeon 5450 running
Linux{\copyright} in 64-bit mode. All code is C++,
single-threaded and was compiled using GCC 4.1.2 with option ``-O3''. The tool
takes a layout file as input, parses it, chops polygons and places them in
their corresponding tiles. Each tile is then transformed individually.
\ffref{transflow} shows flow diagrams of the different steps involved in the
process of transforming a layout using PCHT, DHT, FCFS and FFT. The transform
is initially performed twice to get the machine into steady state, and then
repeated 10 more times to average out the timing noise.
Both PCHT and DHT, as described in \sref{haar_algo}, were fully
custom-implemented. The FFT was performed using the FFTW3 library
\cite{frigo_design_2005}. FCFS was custom-implemented using the TD-FFT
algorithm for pruned FFTs, which in turn use FFTW3 for the sub-FFTs. For the
discrete transforms, an image of the tile is first created and then fed to the
transform algorithm. The time needed to create the discrete image is added to
the runtime of the discrete transform.

For the evaluation, the algorithms were run on three layers from a 22nm layout
of modest size (0.43mm$\times$0.33mm) containing rectangles and more complex
rectilinear polygons. These layers are M1 and M2, which contain both rectangles
and other polygons, and CA, which contains only rectangles, as
shown in \ffref{layout}. We ran the experiment on squared tiles, where side
lengths were powers of two from 128nm$\times$128nm to 4096nm$\times$4096nm.

This experiment has two distinct goals. The first is to validate the
theoretical complexity derived in \sref{algo} as a predictor of the behavior of
the runtime of the transforms. To that effect, we use the runtime from M1 tiled
in 1024nm$\times$1024nm. We compute the average of the 10 runs for each tile
and for all transforms. We then take the median of this average over all tiles
containing a given number of vertices $K$. This is shown in \ffref{haar_comp} and
\ffref{fourier_comp}. We plot $K$ on the $x$-axis, the median runtime on the
left $y$-axis, and the complexity on the right $y$-axis. The plot also shows in
light gray the empirical distribution of $K$ to indicate which range of $K$ is
the most important one.

The second is to measure relative difference of runtime between PCHT and
FCFS, and their respective discrete counterparts. To that effect, we aggregate
the results to get the average of 10 runtimes over a full layer, for all layers
and all transforms. Our metric is the speed-up, computed by dividing the
average runtime of the discrete transform by the average runtime of the
corresponding continuous transform for a given layer and tile size. This is shown in
\ffref{haar_speedup} and \ffref{fourier_speedup}. We plot the speed-up of the
average runtime for all layers and tile sizes considered.

\subsection{Benchmark Results}

\begin{figure*}[tb]
  \centering
  \centerline{
    \begin{minipage}{\linewidth}
      \footnotesize Runtime [ms] \hfill Complexity [$\times10^6$ Op] \\
      \centerline{\includegraphics[width=\linewidth]{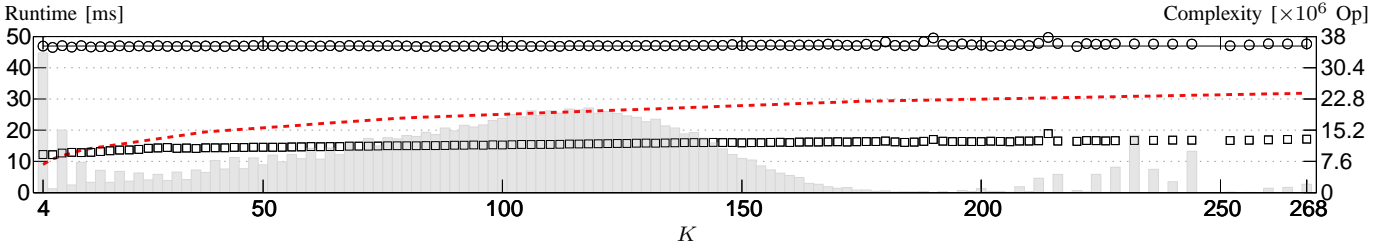}} \\
      \centerline{$K$} \\[-0.4cm]
    \end{minipage}
  }
  \caption{Median runtime (left $y$-axis, data points) and complexity (right $y$-axis,
          lines) of the FCFS ($\square$, dashed line) and FFT ($\circ$, solid line) of
          1024nm$\times$1024nm tiles of the M1 layer containing $K$ vertices.
	  We observe that the gap between the runtimes of the FCFS and the FFT is
          larger than that of their respective complexities.
          The reason being that the pruned FFT can be implemented more efficiently than the FFT
          as shown in \cite{franchetti_generating_2009}.}
  \flabel{fourier_comp}
\end{figure*}

\begin{figure}[tb]
  \centering
  \centerline{
    \begin{minipage}{\linewidth}
      \footnotesize Speed-up \\
      \centerline{\includegraphics[width=\linewidth]{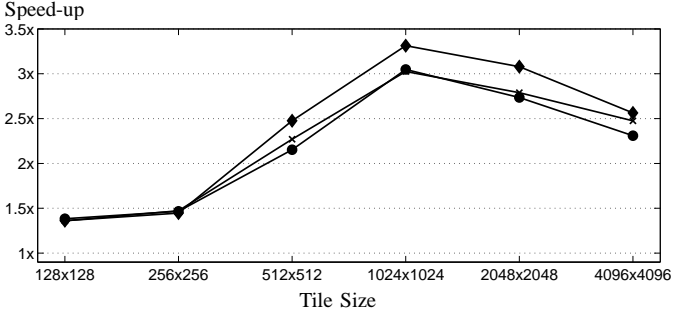}} \\
      \centerline{Tile Size} \\[-0.4cm]
    \end{minipage}
  }
  \caption{Speed-up of the average runtime provided by the FCFS on layers M1 ($\times$),
           M2 ($\blacklozenge$), CA (\mbox{\Large $\bullet$}). The higher the
           better. 1 times speed-up means same performance as the FFT.
           We see that PCHT is at least 1.4 times faster than the FFT. Confidence intervals have been
           omitted in this figure as in the worst case they were found to be 
           within $\pm0.012$ of the value given, with 95\% confidence.}
  \flabel{fourier_speedup}
\end{figure}

\ffref{haar_comp} shows the runtime and the complexity of the PCHT and the DHT
as a function of $K$ for 1024nm$\times$1024nm tiles from M1. The left and right
$y$-axis show the runtime and complexity, respectively. We observe that the
complexity given by \eref{haar_comp} describes the qualitative behavior of the
runtime of the PCHT very well, even for the outliers around $K=190$. When
$K>200$ and the tiles are composed exclusively of rectangles, \eref{haar_comp}
underestimates the complexity. In this case, the runtime is dominated by memory
transfers.

The gap between the runtime of the DHT and its complexity is also explained by
the domination of memory transfers, which are not accounted for in the
computational complexity in \eref{haar_comp}. The dependence of the DHT on $K$
stems from discrete image creation and from the fact that our implementation
uses an \emph{if} statement to avoid storing zero transform coefficients whose
number decreases with increasing $K$. Given the very high number of zero
coefficients, the cost of the \emph{if} statement is justified by the large
memory transfer savings.  Overall, this figure shows that for small values of
$K$ encountered in VLSI layouts, the PCHT is significantly faster than the DHT.

\ffref{haar_speedup} shows that the average runtime of the PCHT for a full M1
layer is about 5 times shorter than that of the DHT, for all considered tile
sizes. For the CA layer, which contains only rectangles that have a lower
complexity, the speed-up is 25 fold for large tiles.

\ffref{fourier_comp} shows the runtime and the complexity of FCFS and the FFT
as a function of $K$ for 1024nm$\times$1024nm tiles from M1. The left and right
$y$-axis show the runtime and complexity, respectively. The runtime of FCFS
compared with that of the FFT is lower than expected given the theoretical
complexity. Indeed, as shown by Franchetti and P\"uschel
\cite{franchetti_generating_2009}, the pruned FFT can be implemented more
efficiently than the FFT. Here again, the FCFS is significantly faster than the
FFT for small values of $K$ found in VLSI layouts.

The speed-up achieved by FCFS over the FFT, shown in \ffref{fourier_speedup},
is similar for all considered layers. The M2 layer shows a slightly higher
speed-up because its vertex density is lower than that of other layers. For all
layers, the highest speed-up found is for 1024nm$\times$1024nm tiles, for which
the FCFS is about 3 times faster than the FFT. In contrast, the results for
128nm$\times$128nm and 256nm$\times$256nm show only a modest speed-up of about
1.5.

For all speed-up results, confidence intervals were computed at the 95\% level
but found to be negligible and have thus been omitted in the figures. However,
they can be found in \tref{perf}, which contains average runtime values of all
considered transforms, for the M1 layer divided into 1024nm$\times$1024nm
tiles, both per tile and for the whole layer. For the 127544 tiles of M1, the
runtime is found to be on average about 5 times lower for the PCHT and 3 times
lower for the FCFS, respectively, than for their discrete counterparts.
Although these numbers might at first glance seem modest, the runtime is reduced from about
9min for the DHT to only 2 using PCHT, whereas the FCFS runs in 30min instead
of 1h40min for the FFT. These are significant time savings.

\begin{table}[t]
\caption{Average runtime of the transforms of an M1 layer divided into
         1024nm$\times$1024nm tiles. There are 127544 non-empty
         tiles for a total of 995497 rectangles and 428817 other polygons.
         Runtime per tile is in $\mu$-seconds. Total runtime is in seconds.}
\tlabel{perf}
\centering
\begin{tabular}{@{}lr@{}lr@{}lr@{}lr@{}l@{}}
\toprule
\multicolumn{9}{@{}l@{}}{\textbf{Average runtime} (with 95\% confidence intervals)} \\
\midrule
       & \multicolumn{2}{@{}c@{}}{\scriptsize PCHT} & \multicolumn{2}{@{}c@{}}{\scriptsize DHT} & \multicolumn{2}{@{}c@{}}{\scriptsize FCFS} & \multicolumn{2}{@{}c@{}}{\scriptsize FFT}\\
\cmidrule{2-9}
\mbox{\scriptsize Per Tile [$\mu$s]} & $919$   & $\pm 0.6$ & $4375$ & $\pm 12.8$ & $15648$ & $\pm 10$ & $47341$ & $\pm 8$ \\
\mbox{\scriptsize Total [s]}         & $117.2$ & $\pm 0.2$ & $558$  & $\pm 4.5$  & $1996$  & $\pm 4$  & $6038$  & $\pm 3$ \\
\bottomrule
\end{tabular}
\end{table}

\section{Conclusions and Future Work}
\slabel{conclusion}

We developed PCHT and FCFS, two new fast algorithms for the computation of the
continuous Haar transform and continuous Fourier series, respectively, of
patterns of rectilinear polygons, as are typically found in VLSI layouts. We
showed that the sparsity of the polygon description can indeed be exploited by
continuous transforms, resulting in significant speed-up of the transform
coefficients computation. The complexities of the two algorithms were analyzed
and compared with that of their discrete counterparts. They were found to be
lower in complexity when the number of vertices of the polygons is low, as is
the case in VLSI layouts. We validated this analysis by implementing all
algorithms in a computational lithography software and running a performance
evaluation on VLSI layouts. The results not only confirmed the validity of our
complexity analysis, but also showed that continuous transforms can actually be
implemented more efficiently than their discrete counterparts. This is achieved
thanks to a more compact description of the input signal, allowing a better
usage of the available cache and reducing memory transfers. We measured the
gain in performance using a speed-up metric. PCHT was found to run at least 5
times faster than the DHT for all considered layers and tile sizes. A maximum
speed-up of 30 times was achieved in the case of the CA layer divided in
4096nm$\times$4096nm tiles. FCFS showed least improvement for small tiles size
where it is only 1.5 times faster than the FFT. However, it showed a peak
performance for 1024nm$\times$1024nm tiles where it is over 3 times faster than
the FFT. These speed-ups result in significant time savings due to the
magnitude of the problem at hand. We therefore conclude that the PCHT and FCFS
are superior to the DHT and FFT, respectively, for rectilinear polygons in VLSI
layouts.

In lithography, often only low-pass Fourier coefficients are needed, and it
would be of high interest to adapt FCFS to yield only this low-pass spectrum.
For the sake of speed, it is currently common practice to employ decimation
using a crude low-pass filter, such as a Haar scaling function, followed by an
FFT, accepting the inevitable large imprecision. Straightforward adaptation
of FCFS using input and output pruned FFT, or even Goertzel, to yield only the
low-pass unaliased spectrum does not compete in terms of speed with the
aforementioned decimation scheme. More sophisticated spectral methods need thus
to be investigated to find a suitable trade-off between speed and accuracy.

We designed PCHT and FCFS with highly suitable structures for parallelization.
Given the trend towards multi- and manycore architectures, such a parallel
implementation would be a natural next step. Given the close relationship
between the Fourier and cosine transforms, straightforward extension of the
FCFS algorithm to the continuous cosine series is possible. A further research
avenue would be to apply continuous transforms to other application
domains. For example, relaxing the signal model to non-rectilinear polygons or
even radically different sparse parametric signals.



%

\appendices


\section*{Acknowledgment}
The authors thank Markus P\"uschel for providing useful advice on the FFT and
pruned FFT.

\ifCLASSOPTIONcaptionsoff
  \newpage
\fi



\bibliographystyle{IEEEtran}
\bibliography{IEEEabrv,bibliography}

\begin{thebibliography}{10}
\providecommand{\url}[1]{#1}
\csname url@samestyle\endcsname
\providecommand{\newblock}{\relax}
\providecommand{\bibinfo}[2]{#2}
\providecommand{\BIBentrySTDinterwordspacing}{\spaceskip=0pt\relax}
\providecommand{\BIBentryALTinterwordstretchfactor}{4}
\providecommand{\BIBentryALTinterwordspacing}{\spaceskip=\fontdimen2\font plus
\BIBentryALTinterwordstretchfactor\fontdimen3\font minus
  \fontdimen4\font\relax}
\providecommand{\BIBforeignlanguage}[2]{{%
\expandafter\ifx\csname l@#1\endcsname\relax
\typeout{** WARNING: IEEEtran.bst: No hyphenation pattern has been}%
\typeout{** loaded for the language `#1'. Using the pattern for}%
\typeout{** the default language instead.}%
\else
\language=\csname l@#1\endcsname
\fi
#2}}
\providecommand{\BIBdecl}{\relax}
\BIBdecl

\bibitem{moore_cramming_1965}
G.~E. Moore, ``Cramming more components onto integrated circuits,''
  \emph{Electronics}, vol.~38, no.~8, pp. 114--117, Apr. 1965.

\bibitem{mack_fundamental_2008}
C.~Mack, \emph{Fundamental Principles of Optical Lithography: The Science of
  Microfabrication}.\hskip 1em plus 0.5em minus 0.4em\relax Wiley, Jan. 2008.

\bibitem{goodman_introduction_2004}
J.~W. Goodman, \emph{Introduction to {F}ourier Optics}, 3rd~ed.\hskip 1em plus
  0.5em minus 0.4em\relax Roberts \& Company Publishers, Dec. 2004.

\bibitem{kwok-kit_wong_resolution_2001}
A.~K.-K. Wong, \emph{Resolution Enhancement Techniques in Optical
  Lithography}.\hskip 1em plus 0.5em minus 0.4em\relax Bellingham, {WA} {USA}:
  {SPIE} Press, 2001.

\bibitem{schellenberg_resolution_2004}
F.~M. Schellenberg and B.~W. Smith, ``Resolution enhancement technology: {T}he
  past, the present, and extensions for the future,'' in \emph{Proc. {SPIE}},
  vol. 5377, May 2004, pp. 1--20.

\bibitem{stulen_extreme_1999}
R.~H. Stulen and D.~W. Sweeney, ``Extreme ultraviolet lithography,''
  \emph{{IEEE} J. Quantum Electron.}, vol.~35, no.~5, pp. 694--699, May 1999.

\bibitem{rosenbluth_optimum_2001}
A.~E. Rosenbluth, S.~J. Bukofsky, M.~S. Hibbs, K.~Lai, A.~F. Molless, R.~N.
  Singh, A.~K.~K. Wong, and C.~J. Progler, ``Optimum mask and source patterns
  to print a given shape,'' in \emph{Proc. {SPIE}}, vol. 4346, Sep. 2001, pp.
  486--502.

\bibitem{poonawala_opc_2006}
A.~Poonawala, P.~Milanfar, and D.~G. Flagello, ``{OPC} and {PSM} design using
  inverse lithography: {A} nonlinear optimization approach,'' in \emph{Proc.
  {SPIE}}, vol. 6154, Mar. 2006, p. 61543H.

\bibitem{kryszczuk_direct_2010}
K.~Kryszczuk, P.~Hurley, and R.~Sayah, ``Direct printability prediction in
  {VLSI} using features from orthogonal transforms,'' in \emph{Proc. of the
  IAPR International Conference on Pattern Recognition}, Aug. 2010.

\bibitem{international_technology_roadmap_for_semiconductors_2007_2007}
{International Technology Roadmap for Semiconductors}, ``2007 edition,''
  \url{http://www.itrs.net/Links/2007ITRS/2007_Chapters/2007_Lithography.pdf},
  2007.

\bibitem{haslam_two-dimensional_1985}
M.~E. Haslam, J.~F. {McDonald}, D.~C. King, M.~Bourgeois, D.~G.~L. Chow, and
  A.~J. Steckl, ``Two-dimensional {H}aar thinning for data base compaction in
  {F}ourier proximity correction for electron beam lithography,'' \emph{J. Vac.
  Sci. Technol., B: Microelectronics and Nanometer Structures}, vol.~3, no.~1,
  pp. 165--173, Jan. 1985.

\bibitem{ma_generalized_2007}
X.~Ma and G.~R. Arce, ``Generalized inverse lithography methods for
  phase-shifting mask design,'' \emph{Opt. Express}, vol.~15, no.~23, pp.
  15\,066--15\,079, Nov. 2007.

\bibitem{antoine_two-dimensional_2004}
J.~Antoine, R.~Murenzi, P.~Vandergheynst, and S.~T. Ali,
  \emph{{Two-Dimensional} Wavelets and their Relatives}.\hskip 1em plus 0.5em
  minus 0.4em\relax Cambridge University Press, 2004.

\bibitem{scheibler_pruned_2010}
R.~Scheibler, P.~Hurley, and A.~Chebira, ``Pruned continuous {H}aar transform
  of 2{D} polygonal patterns with application to {VLSI} layouts,'' to appear
  in: \emph{The Int. Cong. on Computer Applications and Computational Science},
  2010.

\bibitem{chow_image_1983}
D.~G.~L. Chow, J.~F. {McDonald}, D.~C. King, W.~Smith, K.~Molnar, and A.~J.
  Steckl, ``An image processing approach to fast, efficient proximity
  correction for electron beam lithography,'' \emph{J. Vac. Sci. Technol., B:
  Microelectronics and Nanometer Structures}, vol.~1, pp. 1383--1390, Oct.
  1983.

\bibitem{chen_development_2008}
J.~F. Chen, H.~Liu, T.~Laidig, C.~Zuniga, Y.~Cao, and R.~Socha, ``Development
  of a computational lithography roadmap,'' in \emph{Proc. {SPIE}}, vol. 6924,
  2008, p. 69241C.

\bibitem{lee_fourier_1983}
S.~W. Lee and R.~Mittra, ``{F}ourier transform of a polygonal shape function
  and its application in electromagnetics,'' \emph{{IEEE} Trans. Antennas
  Propag.}, vol.~31, no.~1, pp. 99--103, 1983.

\bibitem{chu_calculation_1989}
F.~L. Chu and C.~F. Huang, ``On the calculation of the {F}ourier transform of a
  polygonal shape function,'' \emph{J. of Physics A: Math. and General},
  vol.~22, pp. L671--L672, 1989.

\bibitem{brandolini_average_1997}
L.~Brandolini, L.~Colzani, and G.~Travaglini, ``Average decay of {F}ourier
  transforms and integer points in polyhedra,'' \emph{Ark. Mat}, vol.~35,
  no.~2, pp. 253--275, 1997.

\bibitem{lu_computable_2009}
Y.~M. Lu, M.~N. Do, and R.~S. Laugesen, ``A computable {F}ourier condition
  generating alias-free sampling lattices,'' \emph{{IEEE} Trans. Signal
  Process.}, vol.~57, no.~5, pp. 1768--1782, 2009.

\bibitem{vetterli_world_2009}
M.~Vetterli, J.~Kova\v{c}evi\'c, and V.~K. Goyal, ``The world of {F}ourier and
  wavelets: Theory, algorithms and applications,''
  \url{http://www.fourierandwavelets.org/}, 2009.

\bibitem{mallat_wavelet_2008}
S.~Mallat, \emph{A Wavelet Tour of Signal Processing: The Sparse Way},
  3rd~ed.\hskip 1em plus 0.5em minus 0.4em\relax Academic Press, Dec. 2008.

\bibitem{ahmed_orthogonal_1975}
N.~U. Ahmed and K.~R. Rao, \emph{Orthogonal Transforms for Digital Signal
  Processing}.\hskip 1em plus 0.5em minus 0.4em\relax {Springer-Verlag} New
  York, Inc., 1975.

\bibitem{duhamel_split_1984}
P.~Duhamel and H.~Hollmann, ``Split radix {FFT} algorithm,'' \emph{Electron.
  Lett}, vol.~20, no.~1, pp. 14--16, 1984.

\bibitem{cooley_algorithm_1965}
J.~W. Cooley and J.~W. Tukey, ``An algorithm for the machine calculation of
  complex {F}ourier series,'' \emph{Math. Comput.}, vol.~19, no.~90, pp.
  297--301, 1965.

\bibitem{goertzel_algorithm_1958}
G.~Goertzel, ``An algorithm for the evaluation of finite trigonometric
  series,'' \emph{The American Mathematical Monthly}, vol.~65, no.~1, pp.
  34--35, Jan. 1958.

\bibitem{sorensen_efficient_1993}
H.~Sorensen and C.~Burrus, ``Efficient computation of the {DFT} with only a
  subset of input or output points,'' \emph{{IEEE} Trans. Signal Process.},
  vol.~41, no.~3, pp. 1184--1200, 1993.

\bibitem{medina-melendrez_input_2009}
M.~{Medina-Melendrez}, M.~{Arias-Estrada}, and A.~Castro, ``Input and/or output
  pruning of composite length {FFTs} using a {DIF-DIT} transform
  decomposition,'' \emph{{IEEE} Trans. Signal Process.}, vol.~57, no.~10, pp.
  4124--4128, 2009.

\bibitem{frigo_design_2005}
M.~Frigo and S.~Johnson, ``The design and implementation of {FFTW3},''
  \emph{Proc. {IEEE}}, vol.~93, no.~2, pp. 216--231, 2005.

\bibitem{franchetti_generating_2009}
F.~Franchetti and M.~P\"uschel, ``Generating high performance pruned {FFT}
  implementations,'' in \emph{{IEEE} Int. Conf. on Acoustics, Speech and Signal
  Processing}, pp. 549--552, 2009.

\end{thebibliography}
\end{document}